\documentclass[traditabstract]{aa} 
\usepackage{graphicx}
\usepackage{txfonts}
\begin{document}
   \title{High braking index pulsar PSR J1640-4631: low-mass neutron star with a large inclination angle?}

   \author{Wen-Cong Chen $^{1,2}$          }

   \institute{$^1$ School of Physics and Electrical Information, Shangqiu Normal University, Shangqiu 476000, China;\\
 $^2$ Department of Physics, University of Oxford, Oxford OX1 3RH, UK;
              \email{chenwc@nju.edu.cn} }

   \date{}


  \abstract
{Recent timing observation constrained the braking index of the X-ray pulsar PSR J1640-4631 to be $n=3.15\pm0.03$, which is the highest value of all pulsars with measured braking indices so far. In this Letter, we investigate whether pulsar braking by combined between the magnetic dipole emission and the gravitational radiation might have a braking index greater than three. For conventional neutron star and low mass quark star candidates, the inferred ellipticities derived by the observed braking index are obviously much larger than the theoretical estimated maximum value. If PSR J1640-4631 is a low-mass neutron star with a mass of $0.1~ \rm M_{\odot}$, the inferred ellipticity can be approximately equal to the theoretical estimated maximum value. Because of the radio-quiet nature of this source, we employ the vacuum gap model developed by Ruderman and Sutherland to constrain the inclination angle to be $87.2-90^{\circ}$. Based on this, we propose that a low-mass neutron star with a large inclination angle can interpret the high braking index and the radio-quiet nature of this source. Future observations such as gravitational wave detection and long-term timing for this source are required to confirm or confute our scenario.
   }

   \maketitle


\section{Introduction}

The source PSR J1640-4631 is an X-ray pulsar discovered by \cite{gott14} in a NuSTAR survey of the Norma region in the Galactic plane. Recently, the timing data offer some important parameters of this source such as the spin frequency $\nu=4.84~\rm s^{-1}$, the frequency derivative $\dot{\nu}=-2.28\times 10^{-11}~\rm s^{-2}$, and the second frequency derivative $\ddot{\nu}=3.38\times 10^{-22}~\rm s^{-3}$ (Archibald
et al. 2016). It is generally thought that the rotational kinetic energy of pulsars is converted into radiation energy through magnetic dipole radiation. From the general power-law of pulsar spin-down $\nu=-K\nu^{n}$ (where $K$ is a constant depending on the momentum of inertia, the magnetic field, and the radius of pulsars), we can obtain the so-called braking index $n=\nu\ddot{\nu}/\dot{\nu}^{2}$. Based on the timing data, \cite{arch16} derived the braking index of PSR J1640-4631 to be $n=3.15\pm0.03$.
The authors performed a series of simulations to identify whether this measurement of the braking index arises from the timing noise. Only 0.01 \% of these simulations yielded a braking index greater than three. Therefore, \cite{arch16} thought that the timing noise indicates a very low level and is very probably not responsible for the measured high braking index.

Until 2015, eight pulsars were known with relatively reliable measurements for the braking indices (Lyne et al. 2015). All of these pulsars have a braking index lower than three, showing that the spin-down mechanism is not pure magnetic dipole radiation.
As a new member of the braking indices population, PSR J1640-4631 challenges the radiation and braking mechanisms of pulsars.
The magnetic field decay can result in a braking index higher than three (Blandford \& Romani 1988; Gourgouliatos \&
Cumming 2015). The change of inclination angle $\alpha$ between the magnetic axis and the spin axis may also lead to a braking index that is different from three. A long-term observation of the Crab pulsar has revealed that its inclination angle $\alpha$  may be slowly changing at a rate $\dot{\alpha}\sim 1^{\circ}\rm ~century^{-1}$ (\cite{lyne13}). Based on the inclination angle change model, $\alpha$ of PSR J1640-4631 was constrained to be $18.5\pm3$ degrees, and is decreasing at a rate of $\dot{\alpha}=- (0.23\pm0.05)^{\circ}\rm ~century^{-1}$ (\cite{eksi16}). According to the wind braking model, \cite{kou16} found that the alignment of the inclination angle can cause the braking index to first increase and then decrease, corresponding to a braking index first higher and then smaller than three. In addition, either a mass quadrupole or a magnetic quadrupole would produce gravitational radiation (\cite{blan88}). The pulsars braking by the magnetic dipole emission and the gravitational radiation could give rise to a braking index of between three to five (Archibald et al. 2016; de Araujo et al. 2016).. Employing the derived ellipticity by the braking index, \cite{de16} concluded that aLIGO would not detect the gravitational wave from PSR J1640-4631, while the planned Einstein Telescope would be able to do so.

We here constrain the properties of PSR J1640-4631.
In Sect. 2 the braking torques caused by the magnetic dipole emission and the gravitational
radiation are applied to investigate the spin-down of this source.
In Sect. 3 we constrain the properties of this source by the inferred ellipticity and the vacuum gap model.
Finally, we summarize the results with a brief conclusion in Sect. 4.

\section{Spin-down of PSR J1640-4631}
We assumed the high braking index of PSR J1640-4631 to originate from the gravitational radiation of a compact object with elastic deformation: its spin-down is dominated by the torques originating
from the magnetic dipole radiation and the gravitational wave emission. The angular momentum loss rate of the pulsar by magnetic dipole radiation
is given by
\begin{equation}
\dot{J}_{\rm md}=-\frac{2B^{2}R^{6}{\rm sin}^{2}\alpha\Omega^{3}}{3c^{3}},
\end{equation}
where $B, R$, and $\Omega$ are the surface magnetic field, the radius, and the spin angular velocity of the
pulsar, respectively. The pure magnetic dipole radiation predicts a braking index of three (Ostriker
\& Gunn 1969).

Considering the gravitational radiation of an axisymmetric pulsar with an ellipticity $\epsilon$, the angular momentum loss rate can be written as (Shapiro \& Teukolsky 1983)
\begin{equation}
\dot{J}_{\rm gr}=-\frac{32GI^{2}\epsilon^{2}\Omega^{5}}{5c^{5}},
\end{equation}
where $G$ is the gravitational constant and $I$ the momentum of inertia of the
pulsar. It is clear that a pure gravitational radiation by a mass quadrupole would lead to a braking
index of five. Since the change rate of angular momentum of the pulsar is $\dot{J}=I\dot{\Omega}=\dot{J}_{\rm md}+\dot{J}_{\rm gr}$,
we can obtain (Palomba 2000)
\begin{equation}
n=\frac{\ddot{\Omega}\Omega}{\dot{\Omega}^{2}}=\frac{3+5r}{1+r},
\end{equation}
where $r=\dot{J}_{\rm gr}/\dot{J}_{\rm md}$ represents the fraction of angular momentum change rate that is due to the gravitational radiation versus the magnetic dipole radiation. In calculation, we ignore the change of the magnetic field, the inclination angle $\alpha$, the mass, and the radius of the pulsar.

The observed data offer the total angular momentum change rate. We can therefore derive the following formula from Eq. (3):
\begin{equation}
\dot{J}_{\rm gr}=\frac{r}{1+r}\dot{J}=\frac{r}{1+r}I\dot{\Omega}.
\end{equation}
Combining Eqs. (2) to (4), we can derive the ellipticity of the pulsar as follows:
\begin{equation}
\epsilon=\sqrt{\frac{(n-3)5c^{5}\dot{\Omega}}{64GI\Omega^{5}}}.
\end{equation}
Taking the canonical momentum of inertia $I_{0}=10^{45}~\rm g\,cm^{2}$ and inserting the observed parameters for PSR J1640-4631, the ellipticity of the pulsar can be constrained to be $4.34 - 5.31\times 10^{-3}$. In Fig. 1 we show the relation between the braking index and the ellipticity for PSR J1640-4631. It is worth noting that only an ellipticity greater than $10^{-3}$ can yield a braking index clearly higher than three.

\begin{figure}
\centering
\includegraphics[width=1.1\columnwidth]{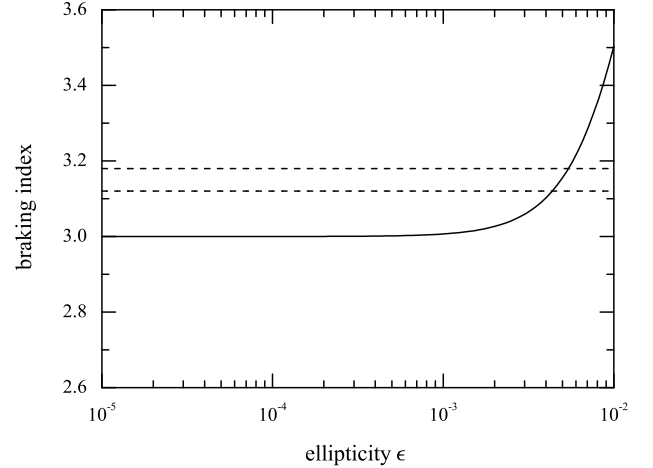}
\caption{The braking index as a function of the ellipticity of the pulsar for a conventional neutron star. Two horizontal dashed lines represent the possible range ($3.12-3.18$) of the measured braking index for the PSR J1640-
4631.}
\label{fig:maps}
\end{figure}

\section{Constraining the properties of PSR J1640-4631}

\subsection{Neutron star}
A conventional neutron star with a fiducial mass, radius, and breaking strain has a maximum ellipticity $\epsilon_{\rm max}=2\times 10^{-7}$ (Owen 2005). Therefore, \cite{arch16} proposed that it can be ruled out that the high braking index is caused by the gravitational radiation. However, the empirical formula for the momentum of inertia given by \cite{bejg02} is invalid for a low-mass neutron star with a mass lower than 0.2 $\rm M_{\odot}$. Theoretical study shows that the lowest mass of a neutron star with a spin frequency 100 Hz is $0.13~\rm M_{\odot}$, and it decreases with decreasing spin frequency (Haensel et al. 2002). Therefore, the lowest mass that might be reached by PSR J1640-4631 at $\nu=4.84$ Hz is $0.1~\rm M_{\odot}$. Observations indicate that low-mass neutron stars may exist in some binary systems such as PSR J1518+4904 ($0.72^{+0.51}_{-0.58}~\rm M_{\odot}$, Janssen et al. 2008), and 4U 1746-37 ($0.41^{+0.70}_{-0.30}~\rm M_{\odot}$ at 99.7 \% confidence, Li et al. 2015). Recently, the calculation performed by \cite{horo10} showed that the highest ellipticity of low-mass ($M = 0.1~\rm M_{\odot}$) neutron stars can reach $5\times 10^{-3}$.  In principle, we should consider the influence of the momentum of inertia in Eq. (5). According to the mass-radius relation of neutron stars $R\propto M^{-1/3}$, a neutron star with $0.1~\rm M_{\odot}$ has a radius $R\simeq 29 ~\rm km$ (we take $R=12$ km for $M=1.4~\rm M_{\odot}$), which leads to a momentum of inertia $I\simeq 0.84\times 10^{45} ~\rm g\,cm^{2}$. Inserting the new $I$ into Eq. (5), the pulsar would need have an ellipticity $\epsilon =5.3\times 10^{-3}$ to yield the high braking index of 3.15. This ellipticity is slightly higher than the theoretical highest ellipticity of a low-mass neutron star. Since the evaluating criterion is not extremely strict, a low-mass neutron star with elastic deformation might be able to produce the strong gravitational radiation and result in the observed braking index.

\subsection{Strange quark star}
Some pulsars have been proposed to consist of strange quarks (Bodmer 1971). Compared with a neutron star, a quark star may have a relatively low mass and small radius (Alcock et al. 1986). \cite{xu05} suggested that the radio-quiet object 1E 1207.4-5209 might be a low-mass bare strange star.  Subsequently, \cite{yue06} argued that if PSR B0943+10 was a low-mass ($0.019~ \rm M_{\odot}$) quark star with a radius of 2.6 km, the problem about its much smaller polar cap area might be solved naturally. Observations have revealed that the Tev luminosity of the pulsar wind nebula HESS J1640-465 is $\sim 6\%$ of spin-down luminosity of a conventional neutron star (Gotthelf et al. 2014). Based on this, we can constrain the lowest momentum of inertia of the pulsar as $I\simeq0.06I_{0}=6\times 10^{43}\rm g\,cm^{2}$.  Because the internal density ($\rho$) in low-mass quark stars is almost homogeneous, its mass can be approximately written as $M=4\pi R^{3}\rho /3$. The density $\rho$ depends on the models of dense matter and is typically a few times the normal nuclear density $\rho_{0}=2.7\times 10^{14}~\rm g\,cm^{-3}$. Based on the lowest momentum of inertia and for $\rho=2\rho_{0}$, the quark star candidate has a mass $M \simeq 0.23~\rm M_{\odot}$,  and the radius of the quark star is $R\simeq5.9~\rm km$.

In the framework of strange quark stars, the highest ellipticity can be written as (Owen 2005)
\begin{equation}
\epsilon_{\rm max}=2\times10^{-4}\frac{\left(\frac{\sigma_{\rm max}}{0.01}\right)\left(\frac{1.4~\rm M_{\odot}}{M}\right)^{2}
\left(\frac{R}{10~\rm km}\right)^{4}}{1+0.14\left(\frac{M}{1.4~\rm M_{\odot}}\right)\left(\frac{10~\rm km}{R}\right)},
\end{equation}
where $\sigma_{\rm max}$ is the breaking strain of the crust. If we take a typical breaking strain of the crust $\sigma_{\rm max}=0.01$, we derive $\epsilon_{\rm max}=8.6\times 10^{-4}$ ($\epsilon_{\rm max}=2.5\times 10^{-3}$ with self-gravity). However, Eq. (5) produces a higher ellipticity $\epsilon\simeq0.02$ according to the momentum of inertia of this quark star $I\simeq6.0\times 10^{43}~ \rm g\,cm^{2}$. The discrepancy is two orders of magnitude between the inferred value from the observation and the theoretical estimated highest value.

\subsection{Radio-quiet nature}
\cite{gott14} discovered PSR J1640-4631 as a pulsating X-ray source in a \emph{NuSTAR} survey. This source is radio quiet (Archibald et al. 2016) and no gamma-ray pulsations have been detected (Gotthelf et al. 2014). The simplest explanation is that the relatively narrow radio beam does not point toward Earth (Brazier \& Johnston 1999). However, its characteristic age is $\tau=-\nu/2\dot{\nu}\approx 3400~\rm yr$, which means that it is a very young pulsar. Young pulsars tend to have a wide radio radiation beam. For example, PSR J1119-6127 (spin period 407.8 ms, characteristic age 1600 yr, both parameters are very similar to those of PSR J1640-4631) has a beam half-opening angle of $\approx30^{\circ}$ (Johnston
\& Weisberg 2006). Another possibility is that this source has already evolved beyond the so-called death line of pulsars (Tr\"{u}mper
2005). In the vacuum gap model, the polar gap surface of the pulsar was thought to be struck by the energetic particles, and is heated and radiates X-rays (Ruderman \& Sutherland 1975). The radio emission originates from the particle acceleration by an effective electric force, which is exerted by the potential drop ($\Phi$) between the magnetic axis and the last open field line. If $\Phi$ is lower than a critical voltage $\Phi_{\rm c}$, the particles cannot be accelerated to an energy high enough to produce sparks. As a result, the pulsar would not give rise to radio emission. The vacuum gap model makes a strong assumption, in which the magnetic axis and the spin axis of the pulsar are aligned. We consider a general case here of an inclination angle $\alpha$ between the magnetic axis and the spin axis. In nonalignment case, the steepest potential drop is given by (Yue et al. 2006)
\begin{equation}
\Phi_{\rm max}\simeq \frac{\Omega BR^{2}}{2c}{\rm cos}\,\alpha({\rm sin}^{2}\theta_{2}-{\rm sin}^{2}\theta_{1}),
\end{equation}
where $\theta_{1}=\alpha-\theta$ (when $\alpha>\theta$), 0 (when $\alpha\leq\theta$), and $\theta_{2}=\alpha+\theta$ (when $\alpha<\pi/2-\theta$), $\pi/2$ (when $\alpha\geq\pi/2-\theta$). The angle $\theta$ is the opening half-angle of the polar cap, and ${\rm sin}\theta =r_{\rm pc}/R$, where $r_{\rm pc}$ is the polar cap radius.

\begin{table}
\begin{center}
\caption{Main parameters for the three candidates of PSR J1640-4631. We list the candidates, the masses, the radius of the pulsar, the polar cap radius, the magnetic fields, and the opening half-angle of the polar cap. \label{tbl-2}}
\begin{tabular}{@{}llllll@{}}
\hline\hline\noalign{\smallskip}
Candidates &$M$               & $R$ &  $r_{\rm pc}$ & $B{\rm sin}\alpha$ &$\theta$  \\
           & ($\rm M_{\odot}$)&(km) & (m)   &  (G) &  (degrees)      \\
\hline\noalign{\smallskip}
CNS  & 1.4   &10   & 319&$1.4\times 10^{13}$& 1.8\\
LMNS & 0.1   &29   & 1575& $4.6\times 10^{11}$&3.2\\
LMQS & 0.23  &5.9  & 145  &$1.7\times 10^{13}$& 2.5 \\
\hline\noalign{\smallskip}
\end{tabular}
\end{center}
\end{table}

To estimate the polar cap radius, we adopt a formula for an aligned pulsar as follows
\begin{equation}
r_{\rm pc}=\sqrt{\frac{2\pi R^{3}}{cP}}.
\end{equation}
The magnetic field of the pulsar can be calculated by
\begin{equation}
B{\rm sin}\alpha\simeq 3.2\times 10^{19}\sqrt{P\dot{P}}\left(\frac{M}{1.4~\rm M_{\odot}}\right)^{1/2}\left(\frac{R}{10~\rm km}\right)^{-2}~\rm G.
\end{equation}
Because $r\approx 0.08$ is very low, we assume a pure magnetic dipole radiation in calculating magnetic field. The two cases are different by a factor $\sqrt{1+r}\approx 1.04$. We consider the influence of inclination angle in calculating magnetic field, while \cite{yue06} assumed $\alpha=90^{\circ}$. In addition, the right side of Eq. (9) is different from that in \cite{yue06} by a factor of two. This difference originates from different magnetic momentum ($\mu$) formulas: we use $\mu=BR^{3}$, while they took $\mu=BR^{3}/2$. To obtain the steepest potential drop, we summarize the related parameters for the candidates of PSR J1640-4631 including conventional neutron star (CNS), low-mass neutron star (LMNS), and low-mass quark star (LMQS) in Table 1.

\begin{figure}
\includegraphics[width=1.1\columnwidth]{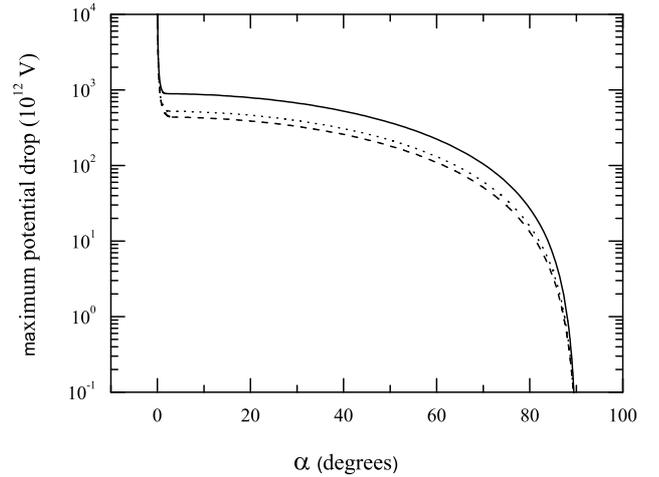}
\caption{Relation between the steepest potential drops and the inclination angle $\alpha$. The solid, dashed, and dotted curves denote the cases of CNS, LMNS, and LMQS, respectively. The input parameters in Eq. (7) are those of Table 1.}
\label{fig:all_model}
\end{figure}

According to Eq. (7), we plot the steepest potential drop $\Phi_{\rm max}$ vs. the inclination angle $\alpha$ in Fig. 2.
The different tendency in the steepest potential drops between our Fig. 2 and Fig. 1 of \cite{yue06} for small inclination angles is noteworthy. This discrepancy is due to including the influence of inclination angle in this work. Taking a critical voltage $\Phi_{\rm c}=10^{12}$ V, nearly all inclination angles (except for large angles of almost $90^{\circ}$) for CNS, LMNS, and LMQS have a potential drop steeper than $\Phi_{\rm c}$. Because it can be ruled out that CNS and LMQS are the candidate of PSR J1640-4631, we can constrain the inclination angle range of LMNS to be $\alpha= 87.2-90^{\circ}$.

\section{Discussion and Summary}

Assuming that the gravitational radiation of a pulsar with elastic deformation can be responsible for the high braking index of PSR J1640-4631, we here constrained its properties. To produce the observed high braking index, the ellipticity of the pulsar is $\epsilon \approx 5\times 10^{-3}$ if the pulsar was a CNS or LMNS. According to the theoretical estimation, it is impossible to possess such a high ellipticity for CNS, while this ellipticity is approximately consistent with the estimated maximum for LMNS. For an LMQS candidate, an ultra-high ellipticity $\epsilon \simeq 0.2$ is required. Based on the vacuum gap model and LMNS candidate, an inclination angle in the range of $\alpha= 87.2-90^{\circ}$ might account for the radio-quiet nature of this source.
Based on the magnetic dipole model with corotating plasma, \cite{eksi16} obtained two possible inclination angles for PSR J1640-4631: $18.5\pm3^{\circ}$ and $56\pm 4^{\circ}$. However, according to the vacuum gap model, CNS can emit in the radio.
If their model is correct, the observer is located exactly in the forbidden zone of the radio emission beam. According to the alignment torque equation given by \cite{phil14}
\begin{equation}
I\frac{{\rm d}\alpha}{{\rm d}t}=-\frac{2B^{2}R^{6}\Omega^{2}}{3c^{3}}{\rm sin}\alpha {\rm cos}\alpha,
\end{equation}
the LMNS model predicts a change rate of the inclination angle $\dot{\alpha}\approx -0.03^{\circ}~\rm century^{-1}$, which is approximately one order of magnitude lower than the value ($-0.23\pm0.05^{\circ}\rm ~century^{-1}$) predicted by \cite{eksi16}. This difference is expected to be explained in the future timing observations.

Perhaps a serious problem in this work is how to yield such an LMNS. In classical stellar evolution theory, it is impossible to form such an LMNS directly through a supernova explosion. An LMNS might be produced through fragmentation during protoneutron star formation or neutron star collisions (Popov et al.
2007; Horowitz 2010). The deformation of an LMNS should be relatively easy to understand because it experiences a violent event. However, the physical mechanism triggering fragmentation is still a puzzle.

To summarize, PSR J1640-4631 cannot be a CNS or LMQS if the high braking index originates from gravitational radiation. However, it is difficult to confirm that this source is an LMNS. One possibility is to try and detect the gravitational wave signals from this source. The characteristic gravitational wave strain equation reads (Abbott et al. 2007)
\begin{equation}
h_{0}=\frac{16\pi^{2} G}{c^{4}}\frac{\epsilon I \nu^{2}}{d},
\end{equation}
where $d$ is the distance of the source. Insertingly, for the observed parameters and the derived ellipticity of PSR J1640-4631
we have $h_{0}=4.2\times 10^{-26}$. In the gravitational wave frequency $\nu_{\rm gw}=9.68 ~\rm Hz$, this signal is lower than the strain sensitivity of aLIGO (de Araujo et al. 2016). Therefore, we expect that the planned Einstein Telescope will investigate this source in the future. According to our scenario, low-mass neutron stars with a braking index greater than three are important gravitational wave sources.

If no gravitational wave from this source is detected, the reasons might be as follows. First, the high braking index of PSR J1640-4631 might be caused by other mechanisms such as the inclination angle change (\cite{eksi16}), the quadrupolar field structure (\cite{petr15}), and the magnetic field decrease. Second, the high braking index may arise from the contaminant of an anomalous $\ddot{\nu}$ after glitches (Alpar \& Baykal 2006). Some pulsars were observed to have an exponential recovery in $\dot{\nu}$ after glitches (see also the panel of PSR J1531-5610 in Fig. 7 of Yu et al. 2013). Such an exponential recovery process in $\dot{\nu}$ could result in a high measured value of $\ddot{\nu}$ (Johnston \& Galloway 1999;
Hobbs et al. 2010), which does not originate from a secular braking torque. Such a glitch might naturally produce an anomalous braking index. Therefore, we expect further observations aimed to detect gravitational waves and long-term timing of this source to confirm or confute our scenario.

\begin{acknowledgements}
We are grateful to the anonymous referee for useful suggestions. We also thank Philipp Podsiadlowski, Aris Karastergiou, R. -X. Xu , X. -D. Li , M. Yu, and  H. Tong for helpful discussions. This work was partly supported by the National Science Foundation of China (under grant number 11573016), the Program for Innovative Research Team (in Science and Technology) at the University of Henan Province, and the China Scholarship
Council.
\end{acknowledgements}

\bibliographystyle{aa}
\bibliography{biblio_NH2CHO}

\end{document}